\documentclass[11pt]{article} 
\usepackage{latexsym}
\usepackage{amssymb}
\usepackage{amsmath}
\usepackage{hyperref}
\hypersetup{colorlinks=true, allcolors=blue}
\usepackage{graphicx}

\usepackage{slashed}
\usepackage{mathbbol}
\usepackage{bm}

\newcommand{\gsim}{\gtrsim}
\newcommand{\tr}{{\rm Tr}}

\usepackage{color}
\usepackage{xcolor}

\usepackage{cite}

\begin{document}
\pagestyle{empty}

\begin{flushright}
KEK-TH-2727\\
YITP-25-123
\end{flushright}

\vspace{3cm}

\begin{center}

{\bf\LARGE
On cusps in the $\eta'$ potential
}
\\

\vspace*{1.5cm}
{\large 
Ryuichiro Kitano$^{1}$,  
Ryutaro Matsudo$^1$ and 
Lukas Treuer$^{1,2,3}$
} \\
\vspace*{0.5cm}

{\it
$^1$Yukawa Institute for Theoretical Physics, Kyoto University, Kyoto 606-8502, Japan,\\
$^2$KEK Theory Center, Tsukuba 305-0801,
Japan\\
$^3$Graduate University for Advanced Studies (Sokendai), Tsukuba
305-0801, Japan\\
}

\end{center}

\vspace*{1.0cm}

\begin{abstract}
{\normalsize 
The large $N$ analysis of QCD states that the potential for the
$\eta'$ meson develops cusps at $\eta' = \pi / N_f$, $3 \pi /N_f$,
$\cdots$, with $N_f$ the number of flavors. Furthermore, the recent discussion of
generalized anomalies tells us that even for finite $N$ there should be
cusps if $N$ and $N_f$ are not coprime, as one can show that
the domain wall configuration of $\eta'$ should support a Chern-Simons
theory on it, i.e., domains are not smoothly connected.
On the other hand, there is a supporting argument for instanton-like, smooth
potentials of $\eta'$ from the analyses of softly-broken
supersymmetric QCD for $N_f= N-1$, $N$, and $N+1$. We argue that the
analysis of the $N_f = N$ case should be subject to the above anomaly
argument, and thus there should be a cusp; while the $N_f = N \pm 1$
cases are consistent, as $N_f$ and $N$ are coprime. We discuss
how this cuspy/smooth transition can be understood.
For $N_f< N$, we find that the number of branches of the $\eta'$ potential is $\operatorname{gcd}(N,N_f)$, which is the minimum number allowed by the anomaly.
We also discuss the condition for s-confinement in QCD-like theories, and find that in general
the anomaly matching of the $\theta$ periodicity indicates that
s-confinement can only be possible when $N_f$ and $N$ are coprime. The s-confinement in
supersymmetric QCD at $N_f = N+1$ is a famous example, and the
argument generalizes for any number of fermions in the adjoint
representation.
}
\end{abstract} 

\newpage
\baselineskip=18pt
\setcounter{page}{2}
\pagestyle{plain}
\baselineskip=18pt
\pagestyle{plain}

\setcounter{footnote}{0}


\section{Introduction}

The QCD in nature, an $SU(3)$ Yang-Mills theory coupled to two or three
flavors of quarks, exhibits rich physics at low energy, such as
confinement and chiral symmetry breaking. While direct
computations of the string tension or particle spectrum are not
easy, we have learned that large $N$ QCD seems to give a
qualitatively good picture where gluon dynamics dominate the
interactions~\cite{tHooft:1973alw}.

The $\eta'$ meson is an interesting particle as it is related to
the chiral anomaly as well as the $\theta$ term in QCD.
The $\eta'$ meson can be treated as a Nambu-Goldstone mode in a
large $N$ limit, while the finite mass can be obtained at the $\mathcal{O}(1/N)$
level. Furthermore, requiring the periodicity $\theta \sim \theta + 2\pi$,
and $1/N$ counting have led to the conclusion that the $\eta'$
potential is not analytic. Instead, it is a simple quadratic form at
the leading order of the $1/N$ expansion. In order to be consistent
with the periodicity, there must be a singularity somewhere between the field values
$\eta' = 0$ and $\eta' = 2\pi / N_f$, where $N_f$ is the
number of flavors~\cite{Witten:1979kh}.
The mass generation at $\mathcal{O}(1/N)$ had explained how the chiral anomaly
gives a mass to the ``Nambu-Goldstone" particle. The discussion also
made it clear that semi-classical instanton analysis does not fully
explain the effects of the anomaly, since instantons are exponentially suppressed at
large $N$.

Although it is convincing that the $\eta'$ meson in real QCD takes the
picture of large $N$, there is no direct proof of it. Indeed, $N=3$ is
not quite large, and $N_f = 2$ or $3$ is not small compared to $N$. It
is thus possible that the dynamics of quarks as important as those of
gluons.
Such a possibility has recently been discussed in
Ref.~\cite{Csaki:2023yas} where the $\eta'$ potential is calculated in
supersymmetric QCD deformed by soft supersymmetry-breaking terms. 
It has been known that softly broken supersymmetric QCD realizes
various phases, including ones resembling
QCD~\cite{Aharony:1995zh, Arkani-Hamed:1998dti, Karch:1998qa,
Luty:1999qc, Abel:2011wv, Kitano:2011zk, Abel:2012un}.
This line of discussion has recently been revived by the findings that
the soft supersymmetry-breaking terms calculated by the anomaly
mediation formula~\cite{Randall:1998uk, Giudice:1998xp} seem to give low-energy physics similar to real QCD~\cite{Murayama:2021xfj,
Csaki:2022cyg}.
Using this technique, it was found that for $N_f = N+1$, $N$ and
$N-1$, the potential is a simple cosine form, which is smooth
everywhere~\cite{Csaki:2023yas}.
Although it is not guaranteed that the $\eta'$ potential they find
continues to be smooth in the non-supersymmetric limit, the analysis
gives an example of the instanton-like picture for the $\eta'$
potential.
This opens up the possibility that our real QCD is of this type, rather
than the large $N$ picture.

In this paper, we discuss the validity of the analysis of
supersymmetric QCD, especially the consistency with the anomaly in the
$\theta$ periodicity~\cite{Cordova:2019uob}.
We argue that there should be a singularity in the
$\eta'$ potential for $N_f = N$, while the smooth potentials for $N_f = N \pm 1$
are consistent.
This means that there is a smooth-cuspy transition when deforming the $N_f = N+1$ theory to $N_f = N$ by turning on a mass term for one of
the quark flavors.
We discuss how such a transition occurs.
For $N_f<N-1$, we find that the number of branches of the $\eta'$ potential is given by the greatest common divisor (gcd) of $N$ and $N_f$, $\operatorname{gcd}(N,N_f)$, which is the minimum number allowed by the anomaly of the $\theta$ periodicity. We argue that there is a $U(1)_{-N+N_f}$ Chern-Simons theory on the domain wall crossing the cusp.
In addition, we discuss the
relation between the anomaly in the $\theta$ periodicity and the
s-confinement phase (confinement without chiral symmetry breaking) of
the gauge theory.
Anomaly matching by massless fermions requires that the greatest common divisor is unity, i.e., $\operatorname{gcd}(N,N_f)=1$.
This is actually a stronger version of the theorem given in
Ref.~\cite{Ciambriello:2024xzd}, where chiral symmetry breaking
should occur for $N_f = p N$, with $p$ an integer.
We also discuss the possibility of testing the $\eta'$ potential on the lattice.

\section{Cusps in the \texorpdfstring{$\eta'$}{eta prime} potential}
\label{sec:cusp}
In Ref.~\cite{Cordova:2019uob}, it has been shown that, in QCD, there
is a mixed anomaly between the vectorial $U(N_f)/\mathbb Z_N$ symmetry and the
$\theta$ periodicity, $\theta \to \theta + 2 \pi$, when gcd$(N, N_f)$
is not unity. This anomaly requires the appearance of a non-trivial theory
on the interface which connects, for example, the $\theta = 0$ and $\theta
= 2 \pi$ regions, as we see below.

In the presence of this mixed anomaly, the $\theta$ periodicity $\theta \sim \theta + 2\pi$
is broken under a topologically non-trivial configuration of the
background gauge fields associated with $U(N_f)/\mathbb Z_N$.
Furthermore, since this is an anomaly, the low-energy effective theory should not
recover the periodicity under the same background configuration, in accordance with 't~Hooft's anomaly matching condition \cite{tHooft:1979rat}.
For example, one can consider a background $U(1)$ baryon gauge
field with Pontryagin index $1/N^2$. Under this background, the
$\theta$ periodicity becomes $\theta \sim \theta + 2 \pi N$, as gluons
are forced to have fractional instanton numbers.
The effective theory for a space-time dependent $\theta$, which is
obtained after integrating out all the dynamical fields, should then
have a structure of
\begin{align}
    -i \int \theta C_4 + i \int_{\rm interface} c_3,
    \label{eq:theta_theory}
\end{align}
in the action. The $C_4$ operator, whose space-time integration should
give $1/N$ modulo $1$ in the background we discussed, cannot be
written in terms of a gauge-invariant combination of background fields
due to the anomaly. Here, gauge invariance includes
the 1-form $\mathbb Z_N$ and $\mathbb Z_{N_f}$
gauge transformations, which are introduced to account for the divisor
part of the global symmetry.
However, one can still define a gauge non-invariant $C_4$, where the gauge
dependence is canceled by a theory on the interface that connects
regions with $\Delta \theta = 2\pi$, i.e., the second term. An example is
\begin{align}
    C_4 = {\frac{N}{8 \pi^2} } \big(B_c^{(2)}\big)^2, {\quad }
    c_3 = - {\frac{1}{4 \pi}} (N c\, dc + 2c\, d \hat A_c),
\end{align}
where $B_c^{(2)}$ and $\hat A_c$ is the pair of $U(1)$ 2-form and $U(1)$ 1-form gauge fields that realizes the 1-form $\mathbb Z_N$ gauge
symmetry, satisfying the relation $N B_c^{(2)} = d \hat A_c$ (see
Ref.~\cite{Gaiotto:2017yup} for the construction); and the 1-form gauge field $c$ is a new
degree of freedom on the interface where the value of $\theta$ changes
rapidly. It is a $U(1)$ 1-form gauge field which forms a $U(1)_{-N}$
Chern-Simons theory on the interface.
The $\mathbb Z_N$ gauge transformation, $\delta B_c^{(2)} = d
\lambda$, $\delta \hat A_c = N \lambda$, and $\delta c = -\lambda$,
with a 1-form parameter $\lambda$, cancels between the two terms.

However, this form is unsatisfactory, as the gauge invariance is
maintained only for configurations with $\Delta \theta = 2\pi$ or
integer multiples thereof. One could replace $\theta$ with $2 \pi [\theta / (2
\pi)]$, where $[\cdots]$ represents the nearest integer, but
there is another requirement that the dependence on a constant $\theta$
should be canceled by an appropriate axial rotation in the limit of massless quarks. A consistent effective theory can thus be constructed by extending the effective action to 
incorporate the $\eta'$ field, which transforms non-linearly under the
anomalous axial $U(1)$ rotation.
To that end, we promote $\theta \to N_f \eta' + \theta$, and the interface of
$\theta$ to a domain wall of the $N_f \eta' + \theta$ configuration,
so that a field redefinition of $\eta'$ can eliminate the constant
part of $\theta$~\cite{Kitano:2020evx}. We thus obtain
\begin{align}
    -2 \pi i \int \left[ 
        {\frac{N_f \eta' + \theta}{2 \pi}}
    \right] C_4 + i \int_{\rm DW} c_3,
    \label{eq:eta'_theory}
\end{align}
such that the partition function is now independent of the constant part of
$\theta$. If we consider an interface where the change of $\theta$ is
rapid enough, for example $\theta_0$ on one side and $\theta_0 + 2\pi$
on the other, we obtain the interface theory without physical
dependence on $\theta_0$ by solving the equation of motion for
$\eta'$. Similarly, on a domain wall configuration of $\eta'$ and
fixing $\theta$ to be constant, the 3d interface theory appears on the
domain wall.

The potential for $\eta'$ should also be written in terms of the combination
with $\theta$ as
\begin{align}
    V (N_f \eta' + \theta).
\end{align}
The Vafa-Witten theorem states that the potential has a minimum at
$N_f \eta' + \theta = 0$~\cite{Vafa:1983tf}, and from the periodicity of
$\theta$ (in the absence of the non-trivial background) it follows that $N_f
\eta' + \theta = 2 \pi n$ with $n$ integers are also minima.
Examples are a simple cosine type potential:
\begin{align}
    V \propto 1 - \cos \left(N_f \eta' + \theta\right),
\end{align}
and a large-$N$ type potential:
\begin{align}
    V \propto \left(N_f \eta' + \theta - 2 \pi \left[
        {\frac{N_f \eta' + \theta}{2 \pi}}
    \right] \right)^2.
\end{align}
The presence of the theory on the interface means that the $\theta$ and
$\theta + 2\pi$ points are not smoothly connected in the potential.
There should be sectors separated by the nearest integer of $(N_f
\eta' + \theta)/(2\pi)$, and the different sectors would define the
domain or the domain wall in Eq.~\eqref{eq:eta'_theory}.
This excludes the possibility of the potential simply being a cosine
form, as in the case that arises, for example, from
semi-classical instanton calculus.

The discussion is quite general: if the effective theory contains the $\eta'$ field, which transforms as $\eta' \to \eta' + 2 \alpha$ under the axial $U(1)$ rotation,
the effective action should be written in terms of the combination 
$N_f \eta' + \theta$. It then follows that there must be multiple sectors separated 
by the nearest integer of $(N_f \eta' + \theta) / (2 \pi)$ for $\operatorname{gcd}(N,N_f) \neq 1$, to be consistent with the violation of the $\theta$ periodicity.
Therefore, the discussion applies, for example, also to softly broken supersymmetric QCD
in the spontaneously chiral symmetry breaking vacuum, which we discuss in more detail later.

Furthermore, we can restrict the number of branches of the $\eta'$
potential by utilizing the anomaly. As discussed in
Ref.~\cite{Cordova:2019uob}, the periodicity of $\theta$ can be
partially recovered up to
$\theta\sim\theta+2\pi\operatorname{gcd}(N,N_f)$
by introducing
counter terms that include the background gauge fields and $\theta$. This means that, even without changing the branch, the value of
$\theta$ or $N_f \eta'$ can be continuously changed from one value to that
added by $2\pi\operatorname{gcd}(N,N_f)$. Note, however, that changing the branch is energetically favorable. Thus, the minimal number of
branches is $\operatorname{gcd}(N,N_f)$, where the minima are at $N_f\eta'+\theta = 2\pi l$ for
$l=0,\ldots,\operatorname{gcd}(N,N_f)-1$, and the point
$l=\operatorname{gcd}(N,N_f)$ is in the same branch as $l=0$. Since
two minima are in the same branch only if they are separated by an
integer multiple of $\operatorname{gcd}(N,N_f)$, the number of
branches must be an integer multiple of $\operatorname{gcd}(N,N_f)$.

\section{\texorpdfstring{$\eta'$}{eta prime} potential in softly broken supersymmetric QCD}

Let us consider supersymmetric QCD with $N_f = N + 1$. According to
the analysis of Ref.~\cite{Csaki:2023yas}, one can realize chiral
symmetry breaking by adding soft supersymmetry breaking terms. In the
low-energy effective theory in terms of mesons and baryons, one can
find the $\eta'$ field which transforms non-linearly under the axial
$U(1)$ transformation.
The potential for $\eta'$ can be explicitly calculated within the
effective theory, where the superpotential for mesons ($M$) and baryons
($B, \bar B$) is written as
\begin{align}
    W = {\frac{B M \bar B - \det M}{\Lambda^{2N - 1}}}.
    \label{eq:n+1}
\end{align}
It is then found that the potential is of cosine type, without a cusp. 
This result is perfectly consistent with the discussion in the
previous section since $\operatorname{gcd}(N, N+1) = 1$.

However, what is puzzling is the case with $N_f = N$. Since $\operatorname{gcd}(N, N_f) \neq
1$, there should be a cusp in the $\eta'$ potential in this theory.  
One can obtain the $N_f = N$ theory through the deformation of the
$N_f = N + 1$ theory by adding a supersymmetric mass term for the
$(N+1)$th flavor of quarks.
By using this deformation, it is found that the $\eta'$ potential in the low-energy effective theory is still smooth~\cite{Csaki:2023yas}.
This analysis is valid for $m \ll \Lambda$, where $m$ and $\Lambda$ are
the supersymmetric mass and the dynamical scale of the theory. It is
known that even for $m \gg \Lambda$, one can reach the correct
effective theory for $N_f = N$ starting from the $N_f = N+1$ theory with the superpotential
Eq.~\eqref{eq:n+1}. The superpotential of the $N_f = N$ effective
theory is
\begin{align}
    W = X \left( \det M - B \bar B - \Lambda^{2N} \right),
    \label{eq:superpotential}
\end{align}
where $X$ is a Lagrange multiplier putting a constraint on the meson
$M$ and the baryon $B$ field values. The Lagrange multiplier field $X$
is then identified as the $M_{N+1, N+1}$ component of the meson in the $N_f
= N+1$ theory. This continuity supports the absence of a phase
transition as we increase $m$ from the regime $m \ll \Lambda$ to $m \gg
\Lambda$.
By treating $X$ as a dynamical field and looking at the $\eta'$
component (the trace part) of the meson field $M$, it is found that
the $\eta'$ potential is smooth.

This result by itself does not contradict anomaly matching.
That is, for any finite value of $m$, the vectorial symmetry of the theory is
$(U(N) \times U(1))/ \mathbb Z_{N+1}$, which is different from $U(N)/
\mathbb Z_N$ in the $N_f = N$ theory. In this case, the discussion of
anomaly matching does not apply, and the $\eta'$ potential can be
smooth. Indeed, the explicit computation of the $\eta'$ potential for
$m \ll \Lambda$ is reliable and should be correct.

However, for $m \gg \Lambda$, we should be able to ignore the
$(N+1)$th quark, and the low-energy dynamics should be described by
those of the $N_f = N$ theory. In this case, the discussion of anomaly matching should apply,
and thus the $\eta'$ potential should exhibit cusps.
We argue that this should be the correct result. Namely, there is no cusp 
for $N_f = N\pm1$ while the cusp appears for $N_f = N$. 
In fact, if we just start from the $N_f = N$ theory, the $X$ field in
Eq.~\eqref{eq:superpotential} is a Lagrange multiplier that is
non-dynamical. With the constraint obtained from the equation of motion
for $X$, one can eliminate the trace part of $M$, i.e., the $\eta'$
meson, from the theory.
Explicitly, in the chiral symmetry breaking vacuum where $\langle M
\rangle = \Lambda^2 {\bf 1}$, we can expand $\det M$ around the vacuum
expectation values as
\begin{align}
    \det M = \Lambda^{2N} e^{\tr \delta M / \Lambda^{2}}
    \left(
    1 - {\frac{\tr (\delta \hat M)^2}{2 \Lambda^4}} - {\frac{(\tr \delta M)^2}{2 N \Lambda^4}}
    + \cdots
    \right),
\label{eq:meson_determinant_expansion}
\end{align}
where we expand $M = \Lambda^2 {\bf 1}+ \delta M$, and $\delta
\hat M$ is the traceless part of $\delta M$, i.e., choosing $\delta M = 1/N\, \tr \delta M\, {\bf 1} + \delta \hat M$.
Here, `$\cdots$' represents
terms of higher order in $\delta M$. The constraint equation
obtained from \eqref{eq:superpotential} is now
\begin{align}
    {\frac{\tr \delta M}{\Lambda^2}}
    = {\frac{\tr (\delta \hat M)^2}{2 \Lambda^4}} 
    + {\frac{B \bar B}{\Lambda^{2N}}} + \cdots,
\label{eq:constrain_equation_traces}
\end{align}
where the $(\tr \delta M)^2$ part of Eq.~\eqref{eq:meson_determinant_expansion} is eliminated by iteratively inserting Eq.~\eqref{eq:constrain_equation_traces} into itself. This constraint can be used to eliminate the trace part of $\delta M$
from the theory without introducing any singularity in the effective action.

Then, the low-energy effective theory does not contain the $\eta'$
field, and thus one cannot calculate the $\eta'$ potential. This is
again consistent with anomaly matching. While there should be a cusp, the
low-energy effective theory just does not see it as $\eta'$ is not a
member of the low-energy degrees of freedom.

From this discussion, we argue that treating $X$ as a dynamical field
is not justified in the $N_f = N$ theory. Similarly, starting from $N_f =
N+1$ and continuing to use the low-energy effective theory for $m \gsim \Lambda$ is
not justified, at least for the discussion of the $\eta'$ potential.

It is also interesting that reliable calculations become possible again for
$N_f = N - 1$ and a smooth $\eta'$ potential is found. This is again
consistent, as gcd$(N, N-1) = 1$. In this theory, the $\eta'$ potential
is indeed obtained from instanton effects.
We will discuss how this series of cuspy-smooth transitions happen in Section~\ref{sec:transitions}.

For $N_f<N-1$, the number of branches of the $\eta'$ potential is
given by $\operatorname{gcd}(N,N_f)$, which is the minimum number of branches allowed by the anomaly of the $\theta$ periodicity, as discussed earlier. In
this case, the number of branches is the number of connected
components of a subspace of the moduli space of vacua in the
supersymmetric limit. Here, we restrict attention to the field space where the pion fields are fixed; otherwise, the entire space would be connected. In this restricted moduli space, the only continuous parameter is the 
R axion, the Nambu–Goldstone boson associated with the spontaneous breaking of the non-anomalous $U(1)_R$ symmetry, which can be
identified with the would-be $\eta'$ meson.
In this subspace, we can have multiple connected components, due to the $\mathbb
Z_{2N_f}$ symmetry, which is the non-anomalous subgroup of the axial
$U(1)$. In the massless limit, the shift of the $\theta$ parameter by $2\pi$ corresponds to the $\mathbb Z_{2N_f}$.  Since two points related by $\mathbb Z_{2N_f}$ can potentially be
connected by a $U(1)_R$ rotation, we need to consider how
many points related by $\mathbb Z_{2N_f}$ are indeed disconnected.
The $U(1)_R$ and $\mathbb Z_{2N_f}$ symmetries act on the gaugino
$\lambda$, the quarks $q,\bar q$ and the scalar quarks $A,\bar A$ as
\begin{align}
  &U(1)_R:\qquad
    \lambda \to e^{-iN_f \alpha/2}\lambda,\quad
    q \to e^{iN\alpha/2}q,\quad \bar q \to e^{iN\alpha/2}\bar q,\quad
    A \to e^{i(N-N_f)\alpha/2}A,
    \notag
    \\
    &\qquad\qquad\quad\bar A \to e^{i(N-N_f)\alpha/2}A,
    \notag
    \\
  &\mathbb Z_{2N_f}:\qquad
  \lambda\to\lambda,\quad
  (q,\bar q, A,\bar A) \to e^{i2\pi/(2N_f)}(q,\bar q,A,\bar A).
  \label{zf}
\end{align}
Note that to simplify the following equations, we here use an unusual convention for the normalization of the $U(1)_R$ charge, where the gaugino does not carry charge unity.
Since the quark condensate vanishes, the moduli space can be parametrized in terms of the gaugino condensate $\lambda\lambda$ and the scalar-quark condensate $\bar AA$.
Furthermore, we restrict attention to the subspace of the moduli space related by $U(1)_R$ and $\mathbb Z_{2N_f}$. In this case, it is sufficient to consider the overall phases of the gaugino and scalar-quark condensates, denoted by the angles $\varphi_\lambda$ and $\varphi_A$, respectively.
Taking $(\varphi_\lambda,\varphi_A) = (0,0)$ as a point at the bottom of the potential, we can parametrize the moduli space by
\begin{align}
  (\varphi_\lambda,\varphi_A) = (-N_f\alpha,(N-N_f)\alpha + 2\pi l/N_f),\quad \alpha\in\mathbb R,\quad l\in\mathbb Z,
  \label{bop}
\end{align}
Let us determine the number of the connected components in
this space, which are labeled by $l$. 
If two components labeled by $l$ and $l+\Delta l$ are the same, the
difference $\Delta l$ is compensated by the change
$\alpha\to\alpha+\Delta\alpha$, which is only possible if both
$-N_f\Delta\alpha\in2\pi\mathbb Z$ and $(N-N_f)\Delta\alpha+2\pi\Delta
l/N_f\in2\pi\mathbb Z$. This implies that $\Delta\alpha = -2\pi J/N_f$ and $\Delta l
= NJ + N_f K$ for some integers $J$ and $K$. Therefore, if the two components are the same, $\Delta l$ is an integer multiple of $\operatorname{gcd}(N,N_f)$.
Conversely, if $\Delta l$ is a multiple of $\operatorname{gcd}(N,N_f)$, the two components are the same, since there exist integers $J,K$ satisfying $NJ+N_fK=\operatorname{gcd}(N,N_f)$ due to B\'{e}zout's identity.
Thus, the different components are labeled by $l=0,1,...,\operatorname{gcd}(N,N_f)-1$, and their number is $\operatorname{gcd}(N,N_f)$.

\section{Check of the anomalies}
We can check that the anomalies of $U(1)_R$ and $\mathbb Z_{2N_f}$ are reproduced in the effective theory in each case.
To that end, let us introduce the background gauge field of the vectorial $U(N_f)/\mathbb Z_N$ symmetry.
Then, the quark fields couple with the gauge field of $G:=[SU(N)\times SU(N_f)\times U(1)]/[\mathbb Z_N\times\mathbb Z_{N_f}]$.
This gauge field is expressed using $su(N)\oplus su(N_f) \oplus u(1)$-valued 1-forms $A^i$ defined on each open subspace $U_i$ of the manifold, and the transition functions $g_{ij}\in G$ defined on $U_{ij}=U_i\cap U_j$, where $\{U_i\}$ is an open cover of the manifold.
The 1-forms $A^i$ and $A^j$ are related by the gauge transformation in terms of $g_{ij}$ on $U_i\cap U_j$.
The transition functions satisfy the cocycle condition $g_{ij}g_{jk}g_{ki} = 1$ on the triple overlaps $U_{ijk} = U_i\cap U_j\cap U_k$, and $g_{ji} = g_{ij}^{-1}$.
We can furthermore express $g_{ij}$ as a triple $(g_{ij}^c, g_{ij}^f,\exp( i\theta_{ij} ))$, where $g_{ij}^c\in SU(N_c)$, $g_{ij}^f\in SU(N_f)$ and $\theta_{ij}\in\mathbb R$; and we have the equivalence relation
$1\sim (1,\exp(2\pi i/N_f),\exp(-2\pi i/N_f)) \sim (\exp(2\pi i/N),1,\exp(-2\pi i/N))$ to account for the $\mathbb Z_N\times\mathbb Z_{N_f}$ part of the quotient group.
Let us then define $g_{ij}^{c,f}$ and $\theta_{ij}$ such that $g_{ji}^{c,f} = ( g_{ij}^{c,f})^{-1}$ and $\theta_{ji} = -\theta_{ij}$.
The cocycle condition and the equivalence relation together imply that
\begin{align}
  &\theta_{ij} + \theta_{jk} + \theta_{ki} = - 2\pi\left(\frac1N b_{ijk} + \frac1{N_f} w_{ijk} \right),\notag\\
  & g_{ij}^cg_{jk}^cg_{ki}^c = e^{2\pi i b_{ijk}/N}\quad g_{ij}^f g_{jk}^f g_{ki}^f = e^{2\pi iw_{ijk}/N_f}, \quad b_{ijk},w_{ijk}\in\mathbb Z.\label{cocycle_theta}
\end{align}
We can identify $\{w_{ijk}\}$ and $\{b_{ijk}\}$ as integer 2-cochains  $w,b\in C^2(M;\mathbb Z)$ by interpreting $U_{ijk}$ as a generator of the chain group $C_2(M)$, where $M$ is the spacetime manifold.
The conditions in Eq.~(\ref{cocycle_theta}) imply that $\delta b = Np$ and $\delta w = -N_f p$ for $p\in C^3(M;\mathbb Z)$. (For a derivation, see Appendix~\ref{app:cocycle_cond}.)
The redundancy in the definition of $b,w$ together with the equivalence relation for $(g_{ij}^c,g_{ij}^f,\exp( i\theta_{ij} ))$, leads to the equivalence relation
\begin{align}
  (b,w)\sim ( b + Nf + \delta h,w-N_f f + \delta g)
  \label{equiv_rel}
\end{align}
for an arbitrary 2-cochain $f$ and arbitrary 1-cochains $g,h$.
Note that only the $su(N_c)$ part of of the 1-form $A^i$ and $g^c_{ij}$ are dynamical, and the other parts, including $b$, are backgrounds.

When the background gauge field couples, the axial $\mathbb Z_{2N_f}$ transformation in Eq.~(\ref{zf}) changes the phase of the partition function by
\begin{align}
  \frac{2\pi}{N_f} \frac1{8\pi^2}\int_M (N_f \tr F_c^2 + N \,\tr F_f^2 + N N_f F_V^2),
\end{align}
where $F_c$, $F_f$ and $F_V$ are the $su(N)$, $su(N_f)$ and $u(1)$ components of the field strength, respectively.
The $su(N)$ part can be rewritten by using \cite{Cordova:2019uob}
\begin{align}
  \frac1{8\pi^2} \int_M F_c^2 = 2\pi \frac{N-1}{2N} \int_M P(b) \bmod 2\pi,
\end{align}
where $P(b)$ is the Pontryagin square \cite{Kapustin:2013qsa,Cordova:2019uob} of $b$, which is defined in terms of cup products as\footnote{
When $N$ is odd, $(N-1)P(b) = (N-1)b\smile b \bmod 2N$ because $N-1$ is even, and $\delta b = Np$. Therefore, the definition is the same as in Ref.\cite{Cordova:2019uob}.
}
$P(b)= b\smile b + b\smile_1 \delta b$, and the integral of a 4-cochain is defined as the value of the 4-cochain evaluated on the fundamental class $[M]$ of $M$.
Then, the change of the phase of the partition function is given by 
\begin{align}
  2\pi \left(\frac{N-1}{2N} \int_M P(b)  + \frac1{8\pi^2 N_f} \int_M (N\,\tr F_f^2 + N N_f F_V^2)\right),
  \label{z2nf_anomaly}
\end{align}
which does not depend on the dynamical field.
One can check that this expression does not violate the equivalence relation (\ref{equiv_rel}).
Under the $U(1)_R$ rotation in Eq.~(\ref{zf}), the partition function changes its phase by
\begin{align}
    N\alpha \frac1{8\pi^2}\int_M (N\,\tr F_f^2 + NN_f  F_V^2).
    \label{u1r_anomaly}
\end{align}

For $N_f<N-1$, in order to reproduce both the $U(1)_R$ and $\mathbb Z_{2N_f}$ anomalies, the effective theory needs to contain the terms
\begin{align}
 &\frac1{8\pi^2}\int_M N\alpha(N\,\tr F_f^2 + NN_f F_V^2) + 2\pi l \left(\frac{N-1}{2N} \int_M P(b)  + \frac1{8\pi^2 N_f} \int_M (N\,\tr F_f^2 + N N_f F_V^2)\right),
 \label{eff_action}
\end{align}
where $\alpha$ and $l$ are the parameters introduced in Eq.~(\ref{bop}).
Several consistency checks for this term are given in Appendix \ref{app:con_check}.
As mentioned in Sec.~\ref{sec:cusp}, there must be dynamical degrees of freedom on the domain wall corresponding to the spatial change of $l$.
For example, if $l$ is 1 in an open region $\tilde M$ and 0 everywhere else, the term 
\begin{align}
  2\pi \frac{N-1}{2N}\int_{\tilde M} P(b)
  \label{pb}
\end{align}
should appear. However, this breaks the equivalence relation $b\sim b + \delta h$ in Eq.~(\ref{equiv_rel}) and is, therefore, ill-defined.

In SUSY QCD with $N_f<N-1$, the $SU(N)$ gauge group is spontaneously broken to $SU(N-N_f)$.
Therefore, the low-energy effective theory consists of the non-linear sigma model of pions, and the pure $SU(N-N_f)$ supersymmetric Yang-Mills theory (SYM).
There are domain walls in this $SU(N-N_f)$ SYM corresponding to $\mathbb Z_{2(N-N_f)}$ symmetry, which acts on the gluino fields and does not change the pion fields.
Under the action of this symmetry, $\alpha$ and $l$ are shifted as $\alpha\to \alpha - 2\pi/(N_f(N-N_f))$ and $l\to l + 1$.
Thus, this domain wall corresponds to the spatial change of $l$.
It was proposed that the surface theory on domain walls separated by $k$ units in $\mathcal N=1$ SYM with gauge group $SU(N)$ at large $N$, is a $U(k)_{-N}$ supersymmetric Chern-Simons theory \cite{Acharya:2001dz}.
Let us assume that this proposal is applicable for finite $N$.
From this, we can conclude that there is a $U(k)_{-N+N_f}$ Chern-Simons theory on the domain wall corresponding to the shift $l\to l+k$. Now, we check whether it can recover the consistency with the equivalence relation (\ref{equiv_rel}).
To that end, let us take $l=k$ on a submanifold $\tilde M$ of $M$, and $l=0$ everywhere else.
The $U(k)_{-N+N_f}$ Chern-Simons theory has a 1-form $\mathbb Z_{N-N_f}$ symmetry, whose action shifts the phase of the Wilson loop,
and the background gauge field for this symmetry should be introduced as an element of the cohomology group $H^2(\tilde M;\mathbb Z_{N-N_f})$.
The cochain $b+w$ can be identified as a representative of an element of $H^2(\tilde M;\mathbb Z_{N-N_f})$, because it satisfies $\delta(b+w) = (N-N_f)p$ and the equivalence relation $b+w \sim b + w + (N-N_f)f + \delta g$.
Thus, we can couple $b+w$ as the background field of the 1-form $\mathbb Z_{N-N_f}$ symmetry of the wall theory.
Since the 1-form $\mathbb Z_{N-N_f}$ symmetry has an 't~Hooft anomaly, this coupling of the background field leads to an inconsistency.
It is known that this inconsistency can be canceled by the 4d term
\begin{align}
  \frac{2\pi k(N-N_f-1)}{2(N-N_f)}\int_{\tilde M} P(w+b).
\end{align}
However, this is different from the term (\ref{pb}), which causes the inconsistency in our effective theory.
To cancel it, we need to add a term that becomes $2\pi$ times an integer when the 4-manifold $\tilde M$ is closed, such that it does not affect anomaly matching for $U(1)_R$ and $\mathbb Z_{2N_f}$.
These terms should be interpreted as 3d counter terms defined on the boundary $\partial \tilde M$.
The explicit form of these terms is given by
\begin{align}
  \frac{2\pi k(N-N_f-1)}{2(N-N_f)}\int_{\tilde M} P(w+b)
   - \frac{2\pi k(N-1)}{2N}\int_{\tilde M} P(b)
  +\frac{2\pi k}{8\pi^2}\int_{\tilde M} \left(\tr F_f^2 +\frac{NN_f}{N-N_f} F_V^2  \right).
  \label{counter_term}
\end{align}
We show that this equals $2\pi$ times an integer when $\tilde M$ is closed in Appendix~\ref{wall_anomaly}.
By adding these terms, we obtain a consistent theory: the first term cancels the inconsistency of the wall theory, the second term is canceled by the term (\ref{pb}), and the last term is consistent on its own.

For $N_f = N+1$, we first consider the supersymmetric limit, followed by the softly broken case.
In the supersymmetric limit, the system is in the s-confinement phase.
The anomaly matching involving $U(1)_R$ was shown in Refs.~\cite{Amati:1988ft,Seiberg:1994bz}, while the one involving $\mathbb Z_{2N_f}$ was shown in Ref.~\cite{Tanizaki:2018wtg}.
For completeness, we confirm the anomaly matching in our notation.
The massless fermions in the effective theory are the fermionic components $\psi_M$, $\psi_B$, $\psi_{\tilde B}$ of the meson superfields $M$ and massless baryon superfields $B$, $\tilde B$, respectively. Their representations for $SU(N_f)_V$, $U(1)_V$, $U(1)_R$ and $\mathbb Z_{2N_f}$ are given by:
\begin{center}
\begin{tabular}{c|c|c|c|c}
& $SU(N_f)_V$ & $U(1)_V$ & $U(1)_R$ & $\mathbb Z_{2N_f}$\\\hline
$\psi_M$ & $\bm{N_f^2-1}$ & $0$ & $ (N-1)/2$ & $2$\\
$\psi_B$ & $\bm{\bar N_f}$ & $N$ & $1/2$ & $N$\\
$\psi_{\tilde B}$  & $\bm{N_f}$ & $N$ & $1/2$&$N$\\
\end{tabular}
\end{center}
Here, we have defined the $U(1)_R$ charges so that a field $\phi_c$ with charge $c$ transforms as $\phi_c\to \exp(ic \alpha)\phi_c$ under the transformation (\ref{zf}).
In this effective theory, the $U(1)_R$ transformation changes the phase of the partition function by
\begin{align}
  \alpha \left( \frac1{8\pi^2}\int_M (N+1)(N-1)\tr F_f^2 + \frac1{8\pi^2}\int_M\left(\tr F_f^2  + (N+1)N^2 F_V^2 \right) \right).
\end{align}
This equals the anomaly (\ref{u1r_anomaly}) of $U(1)_R$ in the original theory.
Under the $\mathbb Z_{2N_f}$ transformation, the phase of the partition function changes by
\begin{align}
  \frac{2\pi}{N+1} \frac1{8\pi^2}\int_M 2(N+1)\tr F_f^2 + \frac{2\pi N}{N+1} \frac1{8\pi^2} \int_M \left(\tr F_f^2 + (N+1)N^2 F_V^2\right).
\end{align}
We can show that this equals the anomaly (\ref{z2nf_anomaly}) by using the relations
\begin{align}
  &-\frac{N-1}{2N}\int_M P(b) + \frac1{8\pi^2}\int_M \left(2\tr F_f^2 + (N-1)N(N+1)F_V^2\right) = 0 \bmod 1.
  \label{s-conf_anomaly}
\end{align}
The derivation of this equation is given in Appendix \ref{app:s-conf_anomaly}.
In the softly broken case, the system is in the chiral symmetry breaking phase.
When we fix the pion degrees of freedom, the moduli space of vacua is parametrized in the same way as Eq.~(\ref{bop}) for $N_f=N+1$.
In this case, there is only one connected component because the shift $l\to l+1$ can be compensated by $\alpha \to \alpha + 2\pi/(N+1)$, which means this space is only parametrized by $\alpha$, the (pseudo) Nambu-Goldstone boson of $U(1)_R$.
Thus, the action of $\mathbb Z_{2N_f}$ equivalent to $l \to l + 1$ expressed in terms of $\alpha$ is given by
\begin{align}
    \alpha \to \alpha - 2\pi/(N+1).
    \label{alpha_change}
\end{align}
In order to reproduce the anomaly (\ref{u1r_anomaly}) of $U(1)_R$, the effective theory contains
\begin{align}
\frac1{8\pi^2}\int_M N\alpha(N\tr F_f^2 + N(N+1)F_V^2).
\end{align}
Let us check that this reproduces the anomaly (\ref{z2nf_anomaly}) of $\mathbb Z_{2N_f}$.
Under the change (\ref{alpha_change}), this term is shifted by
\begin{align}
    -\frac{2\pi}{8\pi^2}\int_M\left(\frac{N^2}{N+1}\tr F_f^2 + N^2 F_V^2\right). 
    \label{shift_n+1}
\end{align} 
To obtain the same form as Eq.~(\ref{z2nf_anomaly}), we use
\begin{align}
\frac{N^2-1}{2N}\int_MP(b) + \frac1{8\pi^2}\int_M(N\tr F_f^2 + N(N+1) F_V^2) = 0\bmod1,
\label{add_to_shift_n+1}
\end{align}
which is the first equation shown in Appendix \ref{app:con_check} for $N_f=N+1$.
By adding $2\pi$ times Eq.~(\ref{add_to_shift_n+1}) to Eq.~(\ref{shift_n+1}), we obtain without changing the phase
\begin{align}
 \text{Eq. }(\ref{shift_n+1}) =
2\pi\frac{N^2-1}{2N}\int_M P(b) + \frac{2\pi}{8\pi^2}\int_M\left(\frac{N}{N+1}\tr F_f^2 + NF_V^2\right) \bmod 2\pi\,.
\label{added_shift_n+1}
\end{align}
Then, we use
\begin{align}
    \frac{N-1}2\int_M P(b) = 0\bmod 1,
    \label{subtract_from_shift_n+1}
\end{align}
which is also shown in Appendix \ref{app:con_check} for a spin manifold $M$, subtracting $2\pi$ times Eq.~(\ref{subtract_from_shift_n+1}) from Eq.~(\ref{added_shift_n+1}). Thus, we find that Eq.~(\ref{shift_n+1}) is equal to the anomaly (\ref{z2nf_anomaly}) of $\mathbb Z_{2N_f}$.

\section{Domain wall transitions}
\label{sec:transitions}

Our claim is that if we start from the $N_f = N+1$ theory and give a
supersymmetric mass $m$ to one of the flavors, then the $\eta'$
potential is smooth for $m \ll \Lambda$ and becomes cuspy for $m \gsim
\Lambda$. Moreover, we will obtain a Chern-Simons theory on the
$\eta'$ domain wall at some critical value $m > m_*$. How does that
happen? It seems strange that a Chern-Simons theory suddenly appears
on the wall as we change the value of $m$ smoothly.
Indeed, it is strongly supported by holomorphy that there is no phase transition
induced by the mass deformation in the superpotential.

To elucidate this, we should first note that in the $N_f = N+1$ theory, the symmetry
breaking pattern is $SU(N + 1)_L \times SU(N + 1)_R \times U(1)_R \to SU(N + 1)_V$ in the vacuum of our interest. Here, $U(1)_R$ is
non-anomalous and the gaugino as well as the (s)quarks are charged under it.
This symmetry is explicitly broken by the gaugino mass, which we take
to be much smaller than the dynamical scale, so that the supersymmetric
results can be used as an approximation.
The low-energy spectrum contains the Nambu-Goldstone fields, i.e., the
pions and the $R$-axion, associated with chiral symmetry
breaking and $U(1)_R$ breaking, respectively.
The $R$-axion, which is a singlet under $SU(N_f)_V$ and called $\eta'$ in
Ref.~\cite{Csaki:2023yas}, is massive due to the explicit (but
small) $U(1)_R$ breaking by the gaugino mass. Thus, since it is light
compared to the dynamical scale, one can reliably use the effective
theory for the calculation of the potential.

However, once we appreciably turn on the mass of one of the flavors of quarks, the symmetry
breaking pattern is now $SU(N)_L \times SU(N)_R \times U(1)_R \to SU(N)_V \times U(1)_R$. Here, the $U(1)_R$ symmetry is different from
the one before. It is rearranged from the original one by adding non-anomalous chiral
charges, so that it remains unbroken by the mass $m$. With the new definition
of $U(1)_R$, the scalar components of the $N \times N$ mesons are not
charged, and thus this $U(1)_R$ is not spontaneously broken in the
vacuum of our interest. Therefore, there is no $R$-axion or $\eta'$ in
the low-energy spectrum, as we already saw in the previous section.

This discussion makes it clear that $\eta'$ in the original $N_f =
N+1$ theory has a mass of order $m$. Similarly, some of the pions in the
$N_f = N+1$ theory also have a mass of order $m$. All those fields have
smooth potentials, and thus there is no Chern-Simons theory on the
domain wall made of them.
Let us consider a massive pion field in the $N_f = N+1$ theory
corresponding to the broken generator $T = {\mbox
{diag}}(1,1,\cdots,1,-N)$. We call
this field $\eta$.
On the other hand, the $R$-axion corresponds to the universal rotation of the quarks, and
we call it $a$. Both fields have masses of $\mathcal{O}(m)$.
We then define $\eta'$ as the linear combination of $\eta$ and $a$ that
corresponds to the generator $T' = {\mbox {diag}}(1,1,\cdots,1,0)$,
where the trace part corresponds to $a$. This field has a mass of
$\mathcal{O}(m)$ for $m \ll \Lambda$, and we expect an $\eta'$ meson to be present
with a mass of $\mathcal{O}(\Lambda)$ for $m \gsim \Lambda$.

Now we define the problem. We consider a domain wall configuration of
the $\eta'$ field whose definition is given above. Specifically, we
take a configuration of $\eta' = 0$ for $z \to -\infty$ and $\eta' =
2\pi/N$ for $z \to \infty$ while all other meson fields are fixed at zero
field values.
This means that this domain wall connects $U=1$ and $U=\operatorname{diag}(e^{2\pi i/N},...,e^{2\pi i/N},1)$, where $U\in U(N+1)$ is the Nambu-Goldstone boson field.
We start with a small value of $m$ ($\ll\Lambda$) where the analysis based on the $N_f=N+1$ effective theory is reliable.
For small $m$, the boundary values of the field $U=1$ and $U=\operatorname{diag}(e^{2\pi i/N},...,e^{2\pi i/N},1)$ are connected by changing only the $\eta$ field.
This is the SUSY QCD version of the domain wall in QCD at $\theta=\pi$.
As discussed in Ref.~\cite{Gaiotto:2017tne}, the theory on the domain wall is a $\mathbb{CP}^{N}$ $\sigma$ model with a Wess-Zumino term.
The appearance of such a non-linear $\sigma$ model can be derived from the bulk theory of Nambu-Goldstone bosons with the background we fixed as above. 
As we change the value of $m$
towards larger values, there should be some critical value of $m_*$
beyond which we obtain a 3d $SU(N)_1$ Chern-Simons theory indicating that the
$\eta'$ potential obtains a cusp.

This phase transition on the domain wall is the same as the one discussed in Ref.~\cite{Gaiotto:2017tne}, where it is argued that this transition can be realized in the $SU(N)_{1-(N+1)/2}$ Chern-Simons theory with $N+1$ fermions.
This transition can also be described by the level-rank dual theory, the $U(1)_{-N}$ Chern-Simons Theory with $N+1$ scalars.
For $m < m^*$, the theory is in the Higgs phase, and thus the low-energy theory is the $\mathbb{CP}^{N}$ $\sigma$ model, while for $m>m_*$, the low-energy effective theory is $U(1)_{-N}$\footnote{
When we consider a domain wall connecting $U=1$ and $U= {\rm diag}(e^{2\pi i k /N}, ..., e^{2\pi i k/N},1)$ with an integer $k \le N$, we can describe it by a $U(k)_{-N}$ theory with $N+1$ scalars. The duality to a fermionic description based on an $SU(N)_{k-(N+1)/2}$ theory was conjectured in Ref.~\cite{Komargodski:2017keh}. 
}.
It is interesting that, while there is no phase transition in the
bulk, the theory on the $\eta'$ domain wall exhibits a phase
transition to separate the $N_f = N+1$ and the $N_f = N$ regimes.

On the other hand, a transition from $N_f = N$ to $N_f = N-1$ is rather trivial. The
$\eta'$ field in this case is associated with the generator $T'' =
{\mbox{diag}(1,1,\cdots,1,-(N-1))}$ of $(SU(N)_L \times SU(N)_R) /
SU(N)_V$. This is one of the pion fields in the $N_f = N$ theory. This field actually does
not obtain a mass of $O(m)$ as
we turn on a mass $m$ for the $N$th flavor, and remains in the spectrum in the $m \gg
\Lambda$ limit. This is because the $U(1)_R$
symmetry breaking is revived. Once we turn on the mass $m$, the definition of the
non-anomalous $U(1)_R$ symmetry is changed such that the massless
meson superfields have a non-vanishing charge.
The symmetry breaking pattern is $SU(N-1)_L \times SU(N-1)_R \times
U(1)_R \to SU(N-1)_V$, and thus $(N-1)^2-1$ pions and an
$R$-axion remain massless up to the supersymmetry-breaking masses. The
nearly massless $R$-axion is called $\eta'$ and has a smooth potential
due to the supersymmetry-breaking terms.
In this case, on an $\eta'$ domain wall, there is no Chern-Simons
theory for any value of $m$.

We could, instead, start the discussion with an $\eta'$ domain wall in
the $N_f = N$ theory rather than that of a pion. In this case, we are
starting with the object outside the low-energy effective theory.
The Chern-Simons theory discussed earlier remains on the domain wall for any value
of $m$, but there is no relation to the domain walls in the low-energy
effective theory.

\section{Condition for the s-confinement in QCD}

The discussion of anomaly matching seems to give non-trivial
constraints on the possible phases of the theory.
Especially, the anomaly makes it difficult to achieve the
s-confinement phase where all the 't~Hooft anomalies are matched by
massless gauge-singlet fermions without chiral symmetry breaking.

In the discussion of the effective theory in Section~\ref{sec:cusp},
it is implicitly assumed that the theory exhibits chiral symmetry
breaking, and there is an $\eta'$ in the spectrum.
However, if we assume that there is no symmetry breaking and try to
reproduce the anomaly in the $\theta$ periodicity only by massless
gauge-singlet fermions, we find that it is impossible for gcd$(N,N_f) \neq
1$. 
For gauge-singlet fermions, the chiral rotation can induce some
background-field-dependent phases, but those are all in the form of
the counter terms in the consideration of the anomaly. The presence of the
anomaly means that $C_4$ in Eq.~\eqref{eq:theta_theory} cannot be
eliminated by those counter terms.
There may be a possibility that something else would reproduce the
anomaly, but at least the standard solutions of anomaly matching do
not work with only massless fermions.
Anomaly matching by massless fermions is possible only for the
theories with gcd$(N,N_f) = 1$, which is, of course, consistent with the
well-known s-confinement phase in supersymmetric QCD with $N_f =
N+1$ as gcd$(N,N+1) = 1$. The discussion applies to general theories
with $U(N_f)/\mathbb Z_N$ global symmetry, such as QCD extended by any
number of adjoint fermions.

Here, it is illustrative to show how the periodicity is matched in the
$N_f = N + 1$ case. In the original, defining theory of gluons and
quarks, the $\theta$ periodicity of the path integral is governed by
the phase factor
\begin{align}
    \exp \left[ 
    i\theta \left(
        Q_c + {\frac{N}{N_f}} Q_f + N Q_V
    \right)
    \right],
    \label{eq:anomaly1}
\end{align} 
where $Q_c$, $Q_f$ and $Q_V$ are the instanton numbers associated with
$SU(N)$, $SU(N_f)_V$ and $U(1)_V$, respectively. They are parametrized by $Q_c
= - m_c m_c' / N + l_c$, $Q_f = - m_f m_f' /N_f + l_f$, and $Q_V =
(m_c / N + m_f / N_f + l_V) (m_c' / N + m_f' /N_f +
l_V')$~\cite{Anber:2019nze}.
The integers $m_c$, $m_c'$, $m_f$, $m_f'$, $l_f$, $l_V$, and $l_V'$ are some
fixed values that depend on the background configurations, and $l_c$
depends on the gluon configuration. For general choices, the
periodicity of $\theta \sim \theta + 2\pi$ is lost due to fractional
instanton numbers.
The term with $Q_c$ is the usual topological term $\propto \theta \int
F \wedge F$. The terms associated with $Q_f$ and $Q_V$ are conventions
and depend on the counter terms. We choose the convention that the
$U(1)_A$ rotation can be compensated by an appropriate shift of
$\theta$. By imposing the same condition for the low-energy theory
with massless mesons and baryons, one finds that the $\theta$ periodicity
is now governed by
\begin{align}
    \exp \left[ 
    i\theta \left(
        \left({\frac{N}{N_f}}+2\right) Q_f + N^3 Q_V
    \right)
    \right].
    \label{eq:anomaly2}
\end{align}
Here, $Q_c$ is absent since the low-energy theory is confined.
Eqs.~\eqref{eq:anomaly1} and \eqref{eq:anomaly2} look totally
different, but their ratio is given by $\exp ({i\theta
\times {\rm integers}})$ for $N_f = N + 1$ under the explicit
substitutions of $Q_c$, $Q_f$ and $Q_V$, and thus the violation of the
periodicity is maintained.
The capability of such an arrangement means that there is no anomaly in
the $\theta$ periodicity in this theory. Conversely, if there is an
anomaly, one cannot arrange massless gauge-singlet fermions to 
maintain the (violation of) periodicity.

\section{Lattice test of the \texorpdfstring{$\eta'$}{eta prime} potential?}

Non-supersymmetric QCD can also be 
directly probed by numerical lattice simulations.
In general, the potential for an operator can be defined as a free
energy while fixing the value of the operator as
\begin{align}
    V(\phi_0) = - {\frac{1}{V_4}} \log Z[\phi_0],
\end{align}
where
\begin{align}
    Z[\phi_0] = \int [d \phi] \delta 
    \left( {\frac{1}{V_4}} \int d^4 x \phi(x) - \phi_0 \right) e^{-S},
    \label{eq:hist}
\end{align}
with $V_4$ being the space-time volume. The lattice simulation evaluates
the path integral as an ensemble average with the probability of
$e^{-S}$. 
With the positivity of the path integral measure, as is the case in
QCD without the $\theta$ term, the partition function $Z[\phi_0]$ can
be interpreted as a histogram of the space-time average of $\phi(x)$,
where the normalization is unimportant as we take the logarithm.

This procedure can, in principle, calculate the potential for fixed
values of $\phi = \int d^4 x\, \bar q q / V_4$. For example, if present, we expect the cusp in the massless limit
at $\eta' = \pi n/N_f$, with $n$ integers.
The projection of the potential minimum on the multi-dimensional $\eta'-\pi$ plane
to the one in the $\phi \propto \cos \eta'$ direction is obtained as $V(\phi)$, which should have a cusp somewhere
if the $\eta'$ potential does.
The difficulty, however, is that we expect a suppression by $e^{-V_4 \Lambda_{\rm QCD}^4}$ for
the probabilities to generate configurations with
$\phi$ deviating from $\langle \phi \rangle$ by $\mathcal{O}(1)$ (in units of $\Lambda_{\rm QCD})$, which is where we would expect a cusp to appear.
Since we eventually need to take the limit of $V_4$ going to infinity to see
a cusp, generating configurations near the cusp
would be statistically not quite easy.

One complication to note is that one cannot identify 
the $\eta'$ potential directly using
the histogram. The quantity,
$\phi_5 \equiv \int d^4 x\, i \bar q \gamma_5 q / V_4 \propto \sin \eta'$, takes pure imaginary values
in the Euclidean path integral
which makes the definition in Eq.~\eqref{eq:hist}
invalid for the $\sin \eta'$ direction.

\begin{figure}
    \centering
    \includegraphics[width=0.5\linewidth]{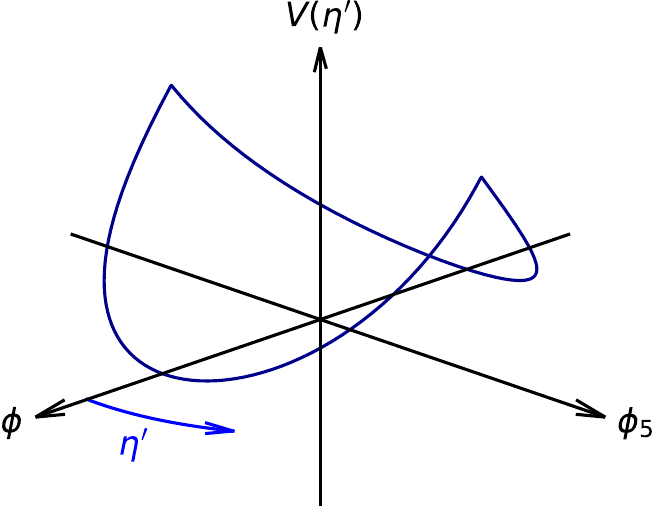}
    \caption{Example for cuspy potential in the $\phi-\phi_5$ plane with $N_f = 2$.}
    \label{fig:phi_phi5_eta_potential}
\end{figure}

Although not easy, in principle,
by introducing imaginary mass terms
to the action, $i \kappa V_4 \phi = \int d^4 x\, i \kappa \bar q q$, we can compute the Fourier transform of the histogram, $\tilde Z(\kappa)$. For the case of $N_f=2$, it is expected that $Z[\phi]$ exhibits a cusp at $\phi=0$, i.e., $\eta' = \pi/2$, implying that the second derivative of $Z[\phi]$ contains a delta-function contribution.
We show an example for the potential for $N_f = 2$ with a cusp at $\phi = 0$ in Fig.~\ref{fig:phi_phi5_eta_potential}. 
If such a delta function is present, the Fourier transform, $\tilde Z(\kappa)$ does not decay to zero in the large $\kappa$ limit. Thus, by analyzing the large-$\kappa$ behavior of $\tilde Z(\kappa)$, one can in principle determine whether a cusp is present in $Z[\phi]$. However, accessing the large-$\kappa$ region requires one to overcome the sign problem, which again poses a significant challenge. 

Although it is an indirect method, testing the qualitative large $N$ behavior such as the Veneziano-Witten formula \cite{Veneziano:1979ec, Witten:1979vv} on the lattice
for various $N_f$ may be
the good starting point towards
understanding the nature of $\eta'$. See for instance Refs.~\cite{Giusti:2001xh, Seiler:2001je, DeGrand:2002gm, Lucini:2004yh, Giusti:2004qd, Luscher:2004fu, DelDebbio:2004ns} for earlier works on the large $N$ $\eta'$ mass, topological susceptibility, and Veneziano-Witten formula, with more recent results found in Refs.~\cite{Bruno:2014ova, Ce:2014sfa, Cichy:2015jra, Verplanke:2024msi, Ottnad:2025zxq, Su:2025ecn}. Furthermore, Ref.~\cite{Nogradi:2019iek} calculates the topological susceptibility for varying $N_f$ with fixed $N = 3$, while Ref.~\cite{DeGrand:2020utq} calculates it for varying $N$ with fixed $N_f = 2$, which are promising steps in the direction of varying $N$ and $N_f$ simultaneously.
See furthermore, e.g., Ref.~\cite{Hernandez:2020tbc} for a recent review on large $N$ QCD on the lattice, including further references.

\section{Conclusions}
Supersymmetric QCD contains an $\eta'$-like particle in the low-energy effective theory when chiral symmetry and $R$-symmetry
are both spontaneously broken. The Nambu-Goldstone particle associated
with the $R$-symmetry, the $R$-axion, is a singlet under
$SU(N_f)_V$, and can be called $\eta'$.
By adding small, soft supersymmetry-breaking terms as a perturbation,
the $\eta'$ field obtains a potential, and it has
been claimed that the potential is given by smooth cosine forms for $N_f = N+1,
N$, and $N-1$.
We have considered the consistency of this analysis confronted with
the anomaly in the periodicity of $\theta$.
We found that, while the $N_f = N+1$ and $N-1$ cases are consistent as
there is no anomaly, the claim that the $\eta'$ potential is smooth for
$N_f = N$ is inconsistent. The correct effective theory of $\eta'$
should exhibit a cusp in the potential so that $\eta'$ and $\eta' +
2\pi/N_f$ are not smoothly connected.

The origin of the discrepancy is that the smooth connection between the $N_f = N+1$ 
and $N_f = N$ regimes does not necessarily imply the absence of 
phase transitions in emergent theories on solitonic objects such as
the Chern-Simons theory on the $\eta'$ domain wall.
Furthermore, the smooth transition of the superpotential does not guarantee that 
the properties of massive modes are also smoothly connected.
In particular, in the effective theory for $N_f = N$, there is no $\eta'$ degree of
freedom as $R$-symmetry is not spontaneously broken, and thus one
cannot discuss the $\eta'$ potential in the effective theory.

It is amusing to note that the effective theory of supersymmetric QCD
is properly evading the $\eta'$ question by not having $\eta'$ in the
spectrum only for the case $N_f = N$ where we have an anomaly.
For generic values of $N$, theories with $N_f < N - 1$ and $N_f > N+1$
are also subject to the anomaly. In this case, the effective theory
either has cuspy $\eta'$ potentials due to multiple vacua of gluino
condensates, or has dual gauge theory descriptions to match the
$\theta$ anomaly.

We thus argue that if the $R$-axion is identified with the $\eta'$, there can be smooth potential
for some $N_f$, while the large $N$ picture works for generic $N$ and
$N_f$. Which one of these captures the actual behavior in real two- or three-
flavor QCD is an open problem. However, our discussion makes it clear
that at least for $N_f = N + 1$, the $R$-axion would not continue to
be the $\eta'$ meson under the mass deformation, and the $N_f = N$
case is in the large $N$ class even for small $N$.

We discussed the transitions among $N_f = N+1, N$ and $N-1$ theories by
deformations from turning on a mass for one of the flavors of quarks.
Between $N_f = N+1$ and $N$, there must be a transition of the 3d
effective theory on the $\eta'$ domain wall from the phase of a $\mathbb{CP}^N$ $\sigma$ model to a Chern-Simons theory.
A natural picture is the realization of a Chern-Simons-Higgs theory
on the domain wall to bridge the two phases of theories.

We have also discussed the relation between the chiral symmetry
breaking in QCD and the anomaly in the $\theta$ periodicity. We find that
s-confinement is only possible for gcd$(N,N_f)=1$ for theories with
the same global symmetry as discussed for QCD; for example, QCD with any number of
fermions in the adjoint representation.

Lastly, although it remains difficult in practice, we briefly discussed possible tests of the $\eta'$ potential using lattice simulations. Unfortunately, a direct test using a histogram for the potential is not easy, since in the infinite-volume limit, we can only calculate the partition function around the minimum of the potential, away from the cusp. Alternative methods pose similar practical challenges, but indirect tests may offer a possible starting point.

\section*{Acknowledgements}
This work is supported in part by JSPS KAKENHI Grant-in-Aid for
Scientific Research (Nos.~19H00689 and 22K21350~[RK, RM]).
This research was supported by the Munich Institute for Astro-, Particle and BioPhysics (MIAPbP) which is funded by the Deutsche Forschungsgemeinschaft (DFG, German Research Foundation) under Germany's Excellence Strategy – EXC-2094 – 390783311.

\appendix

\section{The cocycle conditions for the \texorpdfstring{$U(N_f)/\mathbb Z_N$}{UNf/ZN} gauge field}
\label{app:cocycle_cond}
In this appendix, we show that the cocycle condition (\ref{cocycle_theta}) implies $\delta(Nw+N_f b) = 0$ and $\delta w = N_ff$ for $f\in C^3(M;\mathbb Z)$.
Let us consider four patches $U_i$, $U_j$, $U_k$ and $U_l$ with nonempty overlap $U_i\cap U_j\cap U_k\cap U_l$.
Let $\sigma_{ijkl}$ be a generator of $C_3(M)$ corresponding to the overlap.
The coboundary of $c:= Nw + N_f b$ is written as
\begin{align}
  \delta c(\sigma_{ijkl}) = c_{jkl} - c_{ikl} + c_{ijl} - c_{ijk}.
\end{align}
By using $2\pi c_{ijk}/(NN_f) = \theta_{ij} + \theta_{jk} + \theta_{ki}$ and $\theta_{ji} = -\theta_{ij}$, we can show $\delta c(\sigma_{ijkl})=0$ for any generator $\sigma_{ijkl}$ of $C_3(M)$.
The second equation $\delta w = N_f f$ is rewritten as
\begin{align}
 w_{jkl} - w_{ikl} + w_{ijl} - w_{ijk} = 0 \bmod N_f. \label{cocycle_cond_w}
\end{align}
The third condition in Eq.~(\ref{cocycle_theta}) implies
\begin{align}
  &\exp(2\pi i w_{jkl}/N_f ) =  g^f_{jk}  g^f_{kl}  g^f_{lj}, \notag \\
  &\exp(2\pi i ( w_{ikl} - w_{ijl} + w_{ijk})/N_f) =
  ( g^f_{ij} g^f_{jk} g^f_{ki})
  ( g^f_{ik} g^f_{kl} g^f_{li})
  ( g^f_{ij} g^f_{jl} g^f_{li})^{-1}
  =  g^f_{ij}  g^f_{jk} g^f_{kl}  g^f_{lj}  g^f_{ji}
\end{align}
where we have used $ g^f_{ij} =  (g^f_{ji})^{-1}$.
By taking trace of the above two equations, we obtain
\begin{align}
  &N_f \exp(2\pi i w_{jkl}/N_f ) = \tr ( g^f_{jk}  g^f_{kl}  g^f_{lj}), \notag \\
  &N_f \exp(2\pi i ( w_{ikl} - w_{ijl} + w_{ijk})/N_f) =\tr ( g^f_{jk}  g^f_{kl}  g^f_{lj}).
\end{align}
Since the right-hand sides are equal, it follows that $\exp(2\pi i (w_{jkl}-w_{ikl} + w_{ijl} - w_{ijk})/N_f) = 1$, which implies Eq.~(\ref{cocycle_cond_w}).

\section{Consistency checks of the action of the effective theory}
\label{app:con_check}
In this appendix, we give several consistency checks of the action (\ref{eff_action}) of the effective theory.
Because $l=0$ and $l=N_f$ represent the same point, the actions must be $0\bmod 2\pi$ at $\alpha=0,l=N_f$, i.e.,
\begin{align}
\frac{N_f(N-1)}{2N}\int_MP(b) + \frac1{8\pi^2}\int_M(N\tr F_f^2 + NN_f F_V^2) = 0 \bmod 1.
\label{Nf_wall}
\end{align}
This is shown as follows.
For the integration over a closed manifold, the 1-forms $A_f$ and $A_V$ satisfy
\begin{align}
  &\frac1{8\pi^2}\int \tr F_f^2 = \frac{N_f-1}{2N_f}\int P(w)\  \bmod 1, \qquad\frac1{8\pi^2}\int F_V^2 = \frac12\int(\frac1Nb+\frac1{N_f}w)^2.
  \label{bg_rel}
\end{align}
By using the definition of the Pontryagin square and $\delta b = Np,\delta w=-N_fp$, we obtain
\begin{align}
    &\frac{N_f(N-1)}{2N}\int P(b) = \int \left( \frac{N_f(N-1)}{2N} b^2 + \frac{N_f(N-1)}2 b\smile_1 p\right),\notag\\
    &\frac N{8\pi^2}\int\tr F_f^2 = \int\left(\frac{N(N_f-1)}{2N_f} w^2 - \frac{N(N_f-1)}2 w\smile_1 p\right)\quad \bmod 1,\notag\\
    &\frac{NN_f}{8\pi^2} F_V^2 = \int\left(\frac{N_f}{2N}b^2 + \frac{N}{2N_f} w^2 + \frac12 (bw + wb)\right).
\label{ext_eq}
\end{align}
Because $\delta(N_f b + Nw) = 0$, the mod 2 reduction of $N_f b+Nw$ can be regarded as a representative of an element of $H^2(M;\mathbb Z_2)$.
On a 4 dimensional spin manifold, it is satisfied that $a^2 = 0$ for any $a\in H^2(M;\mathbb Z_2)$. Thus we have
\begin{align}
    0 &= -\frac12\int (N_fb+Nw)^2\quad \bmod 1\notag\\
    &=-\frac12\int(N_f^2 b^2 + N^2 w^2 + NN_f (bw + wb))\quad \bmod 1,\notag\\
    &= -\frac12\int(N_f b^2 + N w^2 + NN_f(bw+wb))\quad \bmod 1,
\label{mod2eq}
\end{align}
where we add $-N_f(N_f-1) \int b^2/2-N(N-1)\int w^2/2\in\mathbb Z$ in the last equation.
By using the property $\delta(u\smile_1 v) = -\delta u\smile_1 v -u\smile_1 \delta v + uv - vu$ of the cup-1 product \cite{Mosher:2008}, we obtain
\begin{align}
    0 &= \frac{NN_f-1}2\int\delta\left(b\smile_1 w + \frac N{N_f}w\smile_1 w\right)\notag\\
    &=\frac{NN_f-1}2\int\left( N_f b\smile_1 p + N w\smile_1 p +  bw -  wb\right)\notag\\
    &=\int\left( \frac{NN_f-N_f}2 b\smile_1 p + \frac{NN_f-N}2 w\smile_1 p + \frac{NN_f-1}2(  bw -  wb)\right)\quad \bmod 1,
    \label{cup1eq}
\end{align}
where we add $-NN_f(N_f-1) \int b\smile_1p/2 -NN_f(N-1)\int w\smile_1p/2\in\mathbb Z$ in the last equation.
By summing both sides of Eqs.~(\ref{ext_eq},\ref{mod2eq},\ref{cup1eq}), we obtain Eq.~(\ref{Nf_wall}).

Because $(\alpha,l) = (-2\pi/N_f,N)$ represents the same point as $(\alpha,l)=(0,0)$, the effective action must be zero at the point, i.e.,
\begin{align}
    \frac{N-1}{2}\int P(b) = 0 \bmod 1.
\end{align}
When $N$ is odd, this holds trivially.
When $N$ is even, we have
\begin{align}
    \frac{N-1}{2}\int P(b) = \frac12 \int b^2 \bmod 1.
\end{align}
Since $\delta b = Np$ with even $N$, $b$ can be regarded as a representative of an element of $H^2(M;\mathbb Z_2)$, and thus $\int b^2 /2 = 0\bmod1$ on any 4 dimensional spin manifold.

\section{Eq.~(\ref{counter_term}) is \texorpdfstring{$0\bmod 2\pi$}{0 mod 2 pi} on a closed manifold}
\label{wall_anomaly}
In this appendix, we show that Eq.~(\ref{counter_term}) is $2\pi$ times an integer when $\tilde M$ is closed.
The definition of the Pontryagin square implies
\begin{align}
  &\frac{N-N_f-1}{2(N-N_f)}\int P(b+w) = \frac{N-N_f-1}{2(N-N_f)}\int \left(b^2 + w^2 + bw + wb + (N-N_f)(b+w)\smile_1 p\right), \notag\\
  &-\frac{N-1}{2N}\int P(b) =-\frac{N-1}{2N}\int b^2 - \frac{N-1}2\int b\smile_1 p.
  \label{pontryagin_squares}
\end{align}
By using Eqs.~(\ref{bg_rel}), we obtain
\begin{align}
  &\frac1{8\pi^2}\int \tr F_f^2 =\frac{N_f-1}{2N_f}\int w^2 - \frac{N_f-1}2\int w\smile_1 p \ \bmod 1,\notag\\
  &\frac{NN_f}{N-N_f}\frac1{8\pi^2}\int F_V^2 =\frac{N_f}{2N(N-N_f)}\int b^2 + \frac{N}{2N_f(N-N_f)}\int w^2 + \frac1{2(N-N_f)}\int(bw + wb).
  \label{continuous_fields}
\end{align}
By using the property $\delta(u\smile_1 v) = -\delta u\smile_1 v -u\smile_1 \delta v + uv - vu$ of the cup-1 product \cite{Mosher:2008}, we obtain
\begin{align}
  &0 = \frac12 \int \delta( b\smile_1 w + \frac N{2N_f} w\smile_1 w) = \frac N2\int w\smile_1 p + \frac{N_f}2\int b\smile_1 p + \frac12\int(bw-wb).
  \label{coboundaries}
\end{align}
By summing Eqs.~(\ref{pontryagin_squares},\ref{continuous_fields},\ref{coboundaries}), we obtain
\begin{align}
  \frac{N-N_f-1}{2(N-N_f)}\int P(w+b)
   - \frac{N-1}{2N}\int P(b)
  +\frac{1}{8\pi^2}\int \left(\tr F_f^2 +\frac{NN_f}{N-N_f} F_V^2  \right) = 0 \bmod 1.
\end{align}
Equation~(\ref{counter_term}) is $2\pi k$ times the left-hand side of this equation, and thus the derivation is complete.

\section{Derivation of Eq.~(\ref{s-conf_anomaly})}
\label{app:s-conf_anomaly}
In this appendix, we derive Eq.~(\ref{s-conf_anomaly}).
By using Eqs.~(\ref{bg_rel}) for $N_f=N+1$, we have
\begin{align}
    &\frac1{8\pi^2}\int\left(2\tr F_f^2 + (N-1)N(N+1) F_V^2 \right) \notag\\
    &= \int\left(\frac N2 w^2 + \frac{(N-1)(N+1)}{2N} b^2 + \frac{N-1}2(bw+wb) \right)\quad \bmod 1.
    \label{appeq0}
\end{align}
By using the property $\delta(u\smile_1 v) = -\delta u\smile_1 v -u\smile_1 \delta v + uv - vu$ of the cup-1 product \cite{Mosher:2008}, we obtain
\begin{align}
    0  = \frac{N-1}{2}\int\delta(b\smile_1 w) =  \frac{(N+1)(N-1)}2\int b\smile_1 p +\frac{N-1}2\int(bw + wb)\quad \bmod 1.
    \label{appeq1}
\end{align}
Because $\delta(Nw + N_f b) = 0$, $Nw+N_f b$ can be regarded as a representative of an element of $H^2(M;\mathbb Z_2)$.
On a 4 dimensional spin manifold, any $a\in H^2(M;\mathbb Z_2)$ satisfies $a^2 = 0$. Thus, we have
\begin{align}
    0 &= \frac12\int (Nw+(N+1)b)^2\quad \bmod 1\notag\\
    &=\frac12\int(N^2 w^2 + (N+1)^2 b^2)\quad \bmod 1,
\end{align}
where we have used the fact that either $N$ or $N+1$ is even.
By adding 
\begin{align}
    0 = \frac12\int\left(- N(N-1)w^2 -N(N+1)b^2\right)\quad\bmod 1
\end{align}
to this equation, we obtain
\begin{align}
    0 =\frac12\int (N w^2 + (N+1) b^2) \quad\bmod 1.
    \label{appeq2}
\end{align}
By adding both sides of Eqs.~(\ref{appeq0},\ref{appeq1},\ref{appeq2}) and
\begin{align}
    0 = -\frac{N(N-1)}2\int b\smile_1 p\quad \bmod 1
\end{align}
we find
\begin{align}
    &\frac1{8\pi^2}\int\left(2\tr F_f^2 + (N-1)N(N+1) F_V^2 \right)
    = \frac{N-1}{2N}\int P(b)\quad\bmod 1.
\end{align}
\bibliographystyle{./utphys.bst}
\bibliography{./eta.bib}

\end{document}